\documentstyle[preprint,aps,epsfig]{revtex}
\newcommand{\eq}{\begin{eqnarray}}
\newcommand{\en}{\end{eqnarray}}
\tightenlines
 
\begin{document}

\title{Updated analysis of meson-nucleon sigma terms \\ 
in the perturbative chiral quark model}

\author{T. \, Inoue, 
V. \ E. \ Lyubovitskij, 
Th. \ Gutsche and 
Amand Faessler\vspace*{0.4\baselineskip}}
\address{
Institut f\"ur Theoretische Physik, Universit\"at T\"ubingen, \\
Auf der Morgenstelle 14,  D-72076 T\"ubingen, Germany 
\vspace*{0.3\baselineskip}}

\maketitle
 
\vskip.5cm                                                                

\begin{abstract} 
We present an updated analysis of meson-baryon sigma terms in the 
perturbative chiral quark model, which is based on effective
chiral Lagrangian. The new feature concerns the inclusion of excited
states in the quark propagator. Its influence on meson loops is shown
to lead in particular for the pion-nucleon sigma term to an enhancement
relevant for the current evaluation of this quantity. 
We also determine various flavor combinations of the scalar nucleon 
form factors and their respective low-momentum transfer limits.
\end{abstract}

\section{Introduction}
Meson-nucleon sigma terms provide a direct measure of the scalar quark 
condensates in the nucleon and thereby also constitute an indicator for 
the mechanism of explicit chiral symmetry breaking. Reviews on sigma-term 
physics including recent theoretical and experimental advances and 
a compilation of results can be found in Ref.~\cite{Gasser:1980sb,Reya:gk,Jaffe:1979rq,Gasser:2000wv,Schweitzer:2003sb}.  

The pion-nucleon sigma term, $\sigma_{\pi N}$, is an essential quantity 
for the study of low energy hadron physics. It is defined by
the scalar density operator $\hat{m}(\bar u u + \bar d d)$ averaged over 
the nucleon or equivalently by $\sigma_{\pi N}(t=0)$, 
the scalar form factor of the nucleon at zero momentum transfer squared.  
The canonical value of the $\pi N$ sigma term 
$\sigma_{\pi N} = 45 \pm 8$ MeV~\cite{Gasser:1990ce} was originally 
extracted from a dispersional analysis of $\pi N$ scattering data, while
in addition exploiting the chiral symmetry constraints. In particular,
the value of the sigma-term, $\sigma_{\pi N} = 45 \pm 8$ MeV, has been 
deduced from the analysis of two quantities: 
$\sigma_{\pi N}(t=2M_{\pi}^2) = 60 \pm 8 $ 
MeV, the scalar nucleon form factor at the Cheng-Dashen (CD) 
point $t=2M_\pi^2$, and the difference $\Delta_\sigma = 
\sigma_{\pi N}(2M_{\pi}^2) - \sigma_{\pi N}(0) = 15.2 \pm 0.4$ 
MeV~\cite{Gasser:1990ce} as induced by explicit chiral symmetry breaking.
The value of the scalar form factor at the CD point 
was close to the one based on a different extraction scheme 
(hyperbolic dispersion relations) from pion-nucleon scattering data:
$\sigma_{\pi N}(t=2M_{\pi}^2) = 64 \pm 8$ MeV~\cite{Koch:pu}. 

An accurate analysis of the $\sigma$ term in chiral perturbation 
theory (ChPT)~\cite{Becher:1999he}, relating the value of $\sigma$ 
at the CD point to the chiral limit, yields $\Delta_\sigma = 14.0$ MeV 
$+ 2 M^4 \bar e_2$; $M$ is the pion  mass in the leading 
order of the chiral expansion and $\bar e_2$ is the fourth-order 
renormalized low-energy constant which is supposed to be small. 
A recent calculation of the shift $\Delta_\sigma$ in ChPT using 
an alternative renormalization scheme~\cite{Fuchs:2003kq} gives a similar 
result of $\Delta_\sigma = 16.9$ MeV $+ 2 M^4 \beta$, where 
$\beta = \bar e_2$. The lattice-regularized ChPT predicts 
a somewhat smaller value of $\Delta_\sigma \sim 10.1 - 11$ MeV within 
the variation of the lattice spacing parameter~\cite{Borasoy:2002hz}. 
Therefore, the ChPT~\cite{Becher:1999he} approaches confirm the original 
result of Ref.~\cite{Gasser:1990ce} with a consistent value for
$\Delta_\sigma$ of about 15 MeV.  

During the last few years updated analyses of also new pion-nucleon 
scattering data lead to a dramatic increase in the value of 
$\sigma_{\pi N}(2M_\pi^2)$: $88 \pm 15$ MeV~\cite{Kaufmann:dd}, 
$71 \pm 9$ MeV~\cite{Olsson:1999jt}, $79 \pm 7$ MeV~\cite{Pavan:2001wz}. 
Although there is criticism~\cite{Stahov:2002vs} on the reliability of 
the recent results on $\sigma_{\pi N}(2M_\pi^2)$, these new values raise 
the result for the empirical $\pi N$ sigma term to $59 - 88$ MeV.
The conclusion of Ref.~\cite{Stahov:2002vs} is that the higher values 
for $\sigma_{\pi N}(t=2M_{\pi}^2)$, which were obtained 
in Refs.~\cite{Kaufmann:dd,Olsson:1999jt,Pavan:2001wz}, can be mainly 
related to the $\pi N$ $D$-wave contribution, but its usage is not
consistent with analyticity.   

Theoretical studies tend to predict a value for the $\pi N$ sigma term, 
which is closer to the canonical result of $\sigma_{\pi N} = 45$ MeV. 
The most naive estimate based on the valence quark picture of the nucleon 
yields only about 15 MeV. The inclusion of the quark-antiquark sea of the 
nucleon is essential to enhance the basic valence quark result.
Theoretical models which somehow take the sea into account therefore give 
a larger reasonable size of the sigma term. For a detailed discussion 
of the theoretical results for $\sigma_{\pi N}$ see  
Ref.~\cite{Schweitzer:2003sb}. Closely linked to the problem of the 
pion-nucleon sigma-term are other related quantities like the 
kaon-nucleon sigma terms and the strangeness content of the nucleon. 

In a previous paper~\cite{Lyubovitskij:2000sf} we performed a detailed 
analysis of meson-nucleon sigma-terms in the context of the perturbative 
chiral quark model (PCQM)~\cite{Lyubovitskij:2000sf,Lyubovitskij:2001nm,Pumsa-ard:2003yh,Lyubovitskij:2001fv,Simkovic:2001fy,Lyubovitskij:2002ng,Khosonthongkee}. 
The PCQM is based on the non-linear $\sigma$-model quark-meson Lagrangian 
and includes a phenomenological confinement potential. Baryons are 
considered as bound states of valence quarks surrounded by a cloud of 
pseudoscalar mesons as imposed by chiral symmetry requirements. This 
model was successfully applied to the electromagnetic properties of the 
nucleon~\cite{Lyubovitskij:2001nm,Pumsa-ard:2003yh}, $\pi N$ scattering 
including radiative corrections~\cite{Lyubovitskij:2001fv}, double 
$\beta$-decay processes~\cite{Simkovic:2001fy}, the strange nucleon 
form factors~\cite{Lyubovitskij:2002ng} and the nucleon axial vector 
coupling constant and form-factor~\cite{Khosonthongkee}. 

In Ref.~\cite{Lyubovitskij:2000sf} we reproduced the conventional value 
of $\sigma_{\pi N} = 45 \pm 5$ MeV which is mostly explained by the pion 
cloud. We showed that the valence quarks give a contribution of $13$ MeV, 
while terms due to the kaon and $\eta$-meson cloud are sufficiently 
suppressed. All calculations for loop diagrams were performed with 
a quark propagator where only the ground state contribution, 
corresponding to hadronic nucleon and delta intermediate states, 
was taken into account. 

In the current work, we update our analysis of meson-nucleon sigma-terms 
and include a discussion of the scalar nucleon form factors. As a new 
feature we investigate the saturation effects of the quark propagator by 
low-lying excited states. In particular, we find that the inclusion of 
excited states leads to an increase of the $\sigma_{\pi N}$ term from 
$\sim 45$ MeV to $\sim 55$ MeV. Additionally, we present the calculation 
of quantities which are not discussed in Ref.~\cite{Lyubovitskij:2000sf}, 
like the scalar pion- and kaon-nucleon form factors and their respective 
slopes, and the value of $\sigma_{\pi N}(2M_\pi^2)$. 

In the present paper we proceed as follows. In Sect.II we review the 
basic notions of the PCQM. In Sec.III we present the calculation of the 
scalar nucleon form factors $\sigma_{\pi N}(Q^2)$ and $\sigma_{K N}(Q^2)$.
We discuss their radii, the value $\sigma_{\pi N}(Q^2)$ at the 
Cheng-Dashen point and at zero recoil (sigma-terms). 
We also give a set of predictions for related quantities: the non-strange 
and strange quark condensates, the strangeness content of the nucleon, 
the isovector $KN$ sigma-term  $\sigma_{KN}^{I=1}$ and others. 
Sec.IV contains a summary of our conclusions. 

\section{The perturbative chiral quark model (PCQM)}
The perturbative chiral quark model~~\cite{Lyubovitskij:2000sf,Lyubovitskij:2001nm,Pumsa-ard:2003yh,Lyubovitskij:2001fv,Simkovic:2001fy,Lyubovitskij:2002ng} is based 
on an effective chiral Lagrangian describing the 
valence quarks of baryons as relativistic fermions moving in an external 
field (static potential) $V_{\rm eff}(r)=S(r)+\gamma^0 V(r)$
with $r=|\vec x|$~\cite{Lyubovitskij:2000sf,Lyubovitskij:2001nm}, 
which in the SU(3)-flavor version are supplemented by a cloud of 
Goldstone bosons $(\pi, K, \eta)$: 
\eq\label{L_chir}
{\cal L}_\chi(x)&=&\bar\psi(x)\,[i \not\!\partial\,-\,\gamma^0 V(r)]\,\psi(x) 
\, + \, \frac{F^2}{4} \, {\rm Tr}[\partial^\mu U(x) \, 
\partial^\mu U^\dagger(x)]\\
&-& \bar\psi(x) \, S(r) \, \biggl[ \frac{U(x) \, + U^\dagger(x)}{2} \, + 
\, \gamma^5 \, \frac{U(x) \, -  U^\dagger(x)}{2} \biggr] \, \psi(x)\nonumber 
\en 
where $U(x)$ is the chiral field and $F=88$ MeV is the pion decay constant 
in the chiral limit~\cite{Gasser:1987rb}. We define the chiral field using 
the exponentional parametrization $U = \exp[i\hat{\Phi}/F]$ where 
$\hat\Phi = \sum\limits_{i=1}^{8} \Phi_i \lambda_i = \sum\limits_{P} 
\Phi_P \lambda_P$ is the octet matrix of pseudoscalar mesons with 
$P = \pi^\pm, \pi^0, K^\pm, K^0, \bar K^0, \eta$ . The relations between 
the sets $\{\Phi_P, \lambda_P\}$ and $\{\Phi_i, \lambda_i\}$  
are given in the Appendix.  

Treating Goldstone fields as small fluctuations around the 
three-quark (3q) core we have the linearized effective 
Lagrangian\footnote{In Eq.~(\ref{linearized_LP}) of the 
Appendix the linearized Lagrangian is given in the basis 
of $\{\Phi_P, \lambda_P\}$.}: 
\eq\label{linearized_L}
{\cal L}_{\rm eff}(x) &=&
\bar{\psi}(x) [i \not\!\partial - S(r) - \gamma^0 V(r)]\psi(x) +
\frac{1}{2} \sum\limits_{i=1}^{8} [\partial_\mu \Phi_i(x)]^2  
\nonumber \\
 &-& \bar{\psi}(x) S(r) i \gamma^5 \frac{\hat \Phi (x)}{F}
 \psi(x)+{\cal L}_{\chi SB}(x).
\en
In Eq.~(\ref{linearized_L}) we include the term ${\cal L}_{\chi SB}$ 
which contains the mass contributions both for quarks and mesons and 
explicitly breaks chiral symmetry: 
\begin{equation}
{\cal L}_{\chi SB}(x) = -\bar\psi(x) {\cal M} \psi(x)
- \frac{B}{2} {\rm Tr} [\hat \Phi^2(x)  {\cal M} ]\,.
\end{equation}
Here, ${\cal M}={\rm diag}\{m_u,m_d,m_s\}$
is the mass matrix of current quarks and 
$B=-\langle 0|\bar u u|0 \rangle / F^2$ is the quark condensate constant. 
In the numerical calculations we restrict to the isospin
symmetry limit $m_u=m_d=\hat m$. After diagonalization the meson mass 
term takes the form 
\begin{equation}
\frac{B}{2} \, {\rm Tr} [\hat \Phi^2(x)  {\cal M} ] = 
\frac{1}{2} \sum_P \, M_P^2 \, \Phi_P^2(x) 
\end{equation} 
where $M_P$ is the set of masses of the pseudoscalar mesons. 
We rely on the standard picture of chiral symmetry 
breaking~\cite{Gasser:1982ap} and for the masses of pseudoscalar 
mesons we use the leading term in their chiral expansion, i.e. 
linear in the current quark masses (see, Eq.(\ref{Meson_Masses}) 
in the Appendix). In the isospin limit they are given by 
\eq\label{M_Masses}
M_{\pi}^2=2 \hat m B, \hspace*{.5cm} M_{K}^2=(\hat m + m_s) B,
\hspace*{.5cm} M_{\eta}^2= \frac{2}{3} (\hat m + 2m_s) B.
\en 
The following set of
parameters~\cite{Gasser:1982ap} is chosen in our evaluation
\begin{equation}
\hat m = 7 \;{\rm MeV},\; \frac{m_s}{\hat m}=25,\;
B = \frac{M^2_{\pi^+}}{2 \hat m}=1.4 \;{\rm GeV}.
\end{equation}
Meson masses obtained in Eq.~(\ref{M_Masses}) satisfy the
Gell-Mann-Oakes-Renner and the Gell-Mann-Okubo relation. In addition,
the linearized effective Lagrangian in Eq.~(\ref{linearized_L})
fulfils the PCAC requirement.

We expand the quark field $\psi$ in the basis of potential
eigenstates as
\eq\label{total_psi}
\psi(x) = \sum\limits_\alpha b_\alpha u_\alpha(\vec{x})
\exp(-i{\cal E}_\alpha t) + \sum\limits_\beta
d_\beta^\dagger v_\beta(\vec{x}) \exp(i{\cal E}_\beta t)\, ,
\en
where the sets of quark $\{ u_\alpha \}$ and antiquark $\{ v_\beta \}$
wave functions in orbits $\alpha$ and $\beta$ are solutions of the
Dirac equation with the static potential $V_{\rm eff}(r)$.
The expansion coefficients $b_\alpha$ and $d_\beta^\dagger$ are the
corresponding single quark annihilation and antiquark creation
operators.

We formulate perturbation theory in the expansion parameter 
$\hat\Phi(x)/F \sim 1/\sqrt{N_c}$ and treat finite current 
quark masses perturbatively~\cite{Lyubovitskij:2001nm}. 
The expansion in powers of $\hat\Phi(x)/F$ corresponds to 
an expansion of matrix elements, formulated in momentum space, 
in powers of $p/F$ where $p$ is the three-momentum of the meson field. 
All calculations are performed at one loop or at order of accuracy 
$o(1/F^2, \hat{m}, m_s)$. In the calculation of matrix elements we project 
quark diagrams on the respective baryon states. The baryon states are 
conventionally set up by the product of the ${\rm SU(6)}$ spin-flavor and
${\rm SU(3)_c}$ color wave functions, where the nonrelativistic single
quark spin wave function is replaced by the relativistic solution
$u_\alpha(\vec{x})$ of the Dirac equation
\begin{equation}\label{Dirac_eq}
\left[ -i\gamma^0\vec{\gamma}\cdot\vec{\nabla} + \gamma^0 S(r) + V(r)
- {\cal E}_\alpha \right] u_\alpha(\vec{x})=0,
\end{equation}
where ${\cal E}_\alpha$ is the single-quark energy. For the description 
of baryon properties we use the effective potential $V_{\rm eff}(r)$ 
with a quadratic radial 
dependence~\cite{Lyubovitskij:2000sf,Lyubovitskij:2001nm}:
\eq\label{V_eff}
S(r) = M_1 + c_1 r^2, \hspace*{1cm} V(r) = M_2+ c_2 r^2
\en
with the particular choice
\eq
M_1 = \frac{1 \, - \, 3\rho^2}{2 \, \rho R} , \hspace*{1cm}
M_2 = {\cal E}_0 - \frac{1 \, + \, 3\rho^2}{2 \, \rho R} , \hspace*{1cm}
c_1 \equiv c_2 =  \frac{\rho}{2R^3} .
\en
Here, ${\cal E}_0$ is the single-quark ground-state energy;
$R$ are $\rho$ are parameters related to the ground-state quark wave
function $u_0$:
\eq\label{Gaussian_Ansatz}
u_0(\vec{x}) \, = \, N \, \exp\biggl[-\frac{\vec{x}^{\, 2}}{2R^2}\biggr]
\, \left(
\begin{array}{c}
1\\
i \rho \, \vec{\sigma}\vec{x}/R\\
\end{array}
\right)
\, \chi_s \, \chi_f \, \chi_c,
\en
where $N=[\pi^{3/2} R^3 (1+3\rho^2/2)]^{-1/2}$ is a normalization
constant; $\chi_s$, $\chi_f$, $\chi_c$ are the spin, flavor and color
quark wave function, respectively. The constant part of the
scalar potential $M_1$ can be interpreted as the constituent mass of
the quark, which is simply the displacement of the current quark mass
due to the potential $S(r)$. The parameter $\rho$ is related to the
axial charge $g_A$ of the nucleon calculated in zeroth-order
(or 3q-core) approximation:

\eq\label{ga_rho_match}
g_A=\frac{5}{3}\biggl(1 - \frac{2\rho^2}{1+\frac{3}{2}\rho^2}\biggr) = 
\frac{5}{3}\biggl(1 - \frac{2}{3}(1 - \gamma) \biggr) 
\en 
where 
\eq
\gamma = \frac{1 - \frac{3}{2} \rho^2}{1 + \frac{3}{2} \rho^2}
\en 
is the relativistic reduction factor (the specific value $\gamma = 1$ 
corresponds to the nonrelativistic limit). 
Therefore, $\rho$ can be replaced by $g_A$ using the matching condition
(\ref{ga_rho_match}). The parameter $R$ is related to the charge radius
of the proton in the zeroth-order approximation as
\eq\label{rad_LO}
\langle r^2_E \rangle^P_{LO} = \int d^3 x \, u^\dagger_0 (\vec{x}) \,
\vec{x}^{\, 2} \, u_0(\vec{x}) \, = \, \frac{3R^2}{2} \,
\frac{1 \, + \, \frac{5}{2} \, \rho^2}{1 \, + \, \frac{3}{2} \, \rho^2}.
\en
In our calculations we use the value 
$g_A$=1.25~\cite{Lyubovitskij:2000sf,Lyubovitskij:2001nm}. 
With $\rho $ fixed, we have only one free parameter, that is $R$. 
In the numerical studies $R$ is varied in the region from 0.55 fm to 0.65 
fm, which corresponds to a change of $\langle r^2_E \rangle^P_{LO}$ from 
0.5 to 0.7 fm$^2$. In the current work we use the central value of 
$R=0.6$ fm. 

The expectation value of an operator $\hat A$ is set up as:
\begin{equation}\label{perturb_A}
\langle \hat A \rangle = {}^B \langle \phi_0 |\sum^{\infty}_{n=0} 
\frac{i^n}{n!}\int d^4 x_1 \ldots \int d^4 x_n T[{\cal L}_I (x_1) 
\ldots {\cal L}_I (x_n) \hat A]|\phi_0 \rangle^B_c,
\end{equation}
where the state vector $|\phi_0 \rangle$ corresponds to the unperturbed
three-quark state ($3q$-core).
Superscript $"B"$ in (\ref{perturb_A}) indicates that the
matrix elements have to be projected onto the respective
baryon states, whereas subscript $"c"$ refers to contributions from
connected graph only. ${\cal L}_I (x)$ of Eq.~(\ref{perturb_A})
refers to the linearized quark-meson interaction Lagrangian:
\begin{equation}
{\cal L}_I (x) = -\bar \psi (x) i \gamma^5 \frac{\hat \Phi (x)}{F}
S(r) \psi(x).
\end{equation}
For the evaluation of Eq.(\ref{perturb_A}) we apply Wick's
theorem with the appropriate propagators for quarks and mesons.

For the quark field we use a Feynman
propagator for a fermion in a binding potential with: 
\eq\label{quark_propagator}
i G_\psi(x,y) &=& \langle 0|T\{\psi(x)\bar \psi(y)\}|0 \rangle
\nonumber \\
&=& \theta(x_0-y_0) \sum\limits_{\alpha} u_\alpha(\vec{x})
\bar u_\alpha(\vec{y}) e^{-i{\cal E}_\alpha (x_0-y_0)}
- \theta(y_0-x_0) \sum\limits_{\beta} v_\beta(\vec{x})
\bar v_\beta(\vec{y}) e^{i{\cal E}_\beta (x_0-y_0)} .
\en
In previous applications~\cite{Lyubovitskij:2000sf,Lyubovitskij:2001nm,Lyubovitskij:2001fv,Simkovic:2001fy,Lyubovitskij:2002ng} we restricted
the expansion of the quark propagator to its ground state with:
\eq\label{quark_propagator_ground}
iG_\psi(x,y) \to iG_0(x,y) \doteq u_0(\vec{x}) \, \bar u_0(\vec{y}) \,
e^{-i{\cal E}_0 (x_0-y_0)} \, \theta(x_0-y_0).
\en
Such a truncation can be considered as an additional regularization of 
the quark propagator, where in the case of SU(2)-flavor intermediate 
baryon states in loop-diagrams are restricted to $N$ and $\Delta$. We 
recently updated our approach by including excited quark states in the 
propagator of Eq.~(\ref{quark_propagator}) and analyzed their influence 
on the matrix elements for the $N$-$\Delta$ transitions 
considered~\cite{Pumsa-ard:2003yh}. Following set of excited quark states 
are included: the first $p$-states ($1p_{1/2}$ and $1p_{3/2}$ in the 
non-relativistic notation) and the second excited states ($1d_{3/2}, 
1d_{5/2}$ and $2s_{1/2}$). For the given form of the effective potential 
(\ref{V_eff}) the Dirac equation can be solved analytically. The 
corresponding expressions for the wave functions of the excited quark 
states are given in the Appendix of Ref.~\cite{Pumsa-ard:2003yh}. In the 
current work we also include the effects of the excited states in the 
quark propagator following the original work of 
Ref.~\cite{Pumsa-ard:2003yh}. 

For the meson fields we adopt the free Feynman propagator with
\eq
i\Delta_{PP^\prime}(x-y) = 
\langle 0|T\{\Phi_P(x)\Phi_{P^\prime}(y)\}|0 \rangle = 
\delta_{PP^\prime}\int\frac{d^4k}{(2\pi)^4i}
\frac{\exp[-ik(x-y)]}{M_P^2 - k^2 - i\epsilon}.
\en

\section{Meson-nucleon sigma terms in the PCQM}

The scalar density operators $S_i^{PCQM}$ $(i=u, d, s)$, relevant for 
the calculation of the meson-baryon sigma-terms in the PCQM, are defined 
as the  partial derivatives of the model $\chi SB$ Hamiltonian 
${\cal H}_{\chi SB}=-{\cal L}_{\chi SB}$ at $m_u \not=  m_d$ with respect 
to the current quark mass of i-th flavor $m_i$. After taking the derivative 
we apply the isospin limit with $m_u = m_d = \hat{m}$. Here we obtain 
\begin{eqnarray}
S_i^{PCQM} \doteq \frac{\partial {\cal H}_{\chi SB}}{\partial m_i} 
\, = \, S_i^{val} \, + \, S_i^{sea}, 
\end{eqnarray}
where $S_i^{val}$ is the set of valence-quark operators coinciding with 
the ones obtained from the QCD Hamiltonian 
\begin{eqnarray}\label{S_val}
S_u^{val} = \bar u u, \hspace*{1cm}  
S_d^{val} = \bar d d, \hspace*{1cm}  
S_s^{val} = \bar s s .  
\end{eqnarray}
The set of sea-quark operators $S_i^{sea}$ arises from the pseudoscalar 
meson mass term:   
\begin{eqnarray}\label{S_sea}
S_u^{sea} &=& B \, \biggl\{ \, \pi^+ \, \pi^- \, + \, 
\frac{\pi^0 \, \pi^0}{2}  \, + \, K^+ \, K^- \, + \, 
\frac{\eta^2}{6} \, \biggr\} \,, \\
S_d^{sea} &=& B \, \biggl\{ \, \pi^+ \, \pi^- \, + \, 
\frac{\pi^0 \, \pi^0}{2}  \, + \, K^0 \, \bar K^0 \, + \, 
\frac{\eta^2}{6} \, \biggr\}  \,,\nonumber\\
S_s^{sea} &=& B \, \biggl\{ \, K^+ \, K^- \, + \, K^0 \, \bar K^0 \, + \, 
\frac{2}{3} \, \eta^2  \, \biggr\}. \nonumber 
\end{eqnarray} 
We calculate the scalar nucleon form factors and meson-baryon sigma-terms 
using Eq.~(\ref{perturb_A}) in the isospin limit ($m_u = m_d=\hat m$) and 
at order of accuracy $o(1/F^2, \hat{m}, m_s)$. For technical details 
concerning the perturbative analysis of Eq. (\ref{perturb_A}) see 
Refs.~\cite{Lyubovitskij:2000sf,Lyubovitskij:2001nm}. For example, 
the expression for the scalar form factor $\sigma_{\pi N}(Q^2)$ 
is given as 
\eq\label{piN_pert}
\sigma_{\pi N}(Q^2) &=& 
\hat{m}  \,\, 
 {}^p \! \langle \phi_0| \sum_{n=0}^{2} \frac{i^n}{n!}\! 
 \int \, \delta(t) \, d^4x \, d^4x_1 \, \ldots \, d^4x_n \, 
 e^{-iqx} \, \\
&\times& 
T[{\cal L}_I(x_1) \ldots {\cal L}_I(x_n) \, S_{u+d}^{PCQM}(x) \, ]
 | \phi_0 \rangle^p_c  \nonumber
\en
where we introduced the shorted notation 
\eq
S_{q \pm q^\prime}^{PCQM} = 
S_{q}^{PCQM} \pm S_{q^\prime}^{PCQM} \,. 
\en
As in previous work, for example in the calculation of electromagnetic
nucleon form factors \cite{Lyubovitskij:2001nm}, we restrict our 
kinematics to a specific frame, that is the Breit frame. This constraint
is sufficient to guarantee local gauge invariance concerning the coupling 
of the electromagnetic field (for a detailed discussion see 
Ref.~\cite{Lyubovitskij:2001nm}). The Breit frame is specified as 
follows: the initial momentum of the nucleon is $p = (E, -\vec{q}/2)$, 
the final momentum is $p^\prime = (E, \vec{q}/2)$ and the 4-momentum of 
the external field is $q = (0, \vec{q}\,)$ with $p^\prime = p + q$. The 
space-like momentum transfer squared is given by 
$Q^2 = - q^2 = \vec{q}^{\, 2}$. 

The following diagrams contribute to the scalar nucleon form factors (and 
therefore the sigma-terms) up to the one-loop level: tree level diagram 
(Fig.1a) with the insertion of the valence-quark scalar density 
$S_i^{val}$ into the quark line, the meson cloud (Fig.1b) and meson 
exchange diagrams (Fig.1c) with insertion of the sea-quark scalar density 
$S_i^{sea}$ to the meson line. We do not take into account the diagram 
generated by dressing of the valence-quark scalar density operator by the 
meson field since it is proportional to $\hat{m}/F^2$ or $m_s/F^2$ and, 
therefore, contributes only to the next order of our perturbation expansion. 

The $\pi N$ sigma term 
$\sigma_{\pi N} = \sigma_{\pi N}(0)$ is then obtained as 
\eq
\sigma_{\pi N} \, = \, \sigma_{\pi N}^{val} \, + \, \sigma_{\pi N}^{sea} 
\en
where 
\eq
\sigma_{\pi N}^{val} \, = \, 3 \gamma \hat m 
\en
is due to the valence quarks and 
\eq\label{sigma_piN_sea}
\sigma_{\pi N}^{sea} = 
\!\! \sum_{\Phi=\pi,K,\eta} \sigma_{\pi N}^{\Phi} = 
\!\! \sum_{\Phi=\pi,K,\eta} \!\!\! d_N^{\Phi \, (1)} \, 
\Gamma_{\Phi}^{(1)}~ +    \! \sum_{\Phi=\pi,K,\eta} \!\!\! 
d_N^{\Phi \, (2)} \, \Gamma_{\Phi}^{(2)}
\en 
is the sea-quark contribution. Here, $d_N^{\Phi \, (i)}$ with  
$\Phi = \pi, K$ or $\eta$ are the recoupling coefficients defining 
the partial contributions of the $\pi$, $K$, and $\eta$-meson cloud 
related to the diagrams of Fig.1b (with $i=1$) and 
Fig.1c (with $i=2$): 
\eq
&&d_N^{\pi\, (1)} = \frac{81}{400}, \hspace*{.6cm} 
d_N^{K\, (1)} = \frac{54}{400}, \hspace*{.6cm} 
d_N^{\eta\, (1)} = \frac{9}{400}, \\
&&d_N^{\pi\, (2)} = \frac{90}{400}, \hspace*{.65cm} 
d_N^{K\, (2)} \equiv 0, \hspace*{1.1cm} 
d_N^{\eta\, (2)} = - \frac{6}{400} \,. \nonumber
\en
Analytical expressions for the vertex functions $\Gamma_{\Phi}^{(1)}$ 
and $\Gamma_{\Phi}^{(2)}$, when the quark propagator is truncated to the 
ground state, are given in Ref.~\cite{Lyubovitskij:2000sf}. 
The additional contribution of excited quark states are evaluated as 
laid out in Ref. ~\cite{Pumsa-ard:2003yh}.  
Note that the Feynman-Hellmann (FH) theorem, relating the pion-nucleon 
sigma-term $\sigma_{\pi N}$ to the nucleon mass $m_N$ 
\begin{eqnarray}\label{FHTh}
\sigma_{\pi N} = \hat m \frac{\partial m_N}{\partial \hat m}. 
\end{eqnarray} 
is fulfiled in the present model calculation for any form of the quark 
propagator (truncated to the ground state or with inclusion of excited 
states) as originally proven in Ref. ~\cite{Lyubovitskij:2000sf}. 
The nucleon mass in our approach is given by 
\eq
m_N =  3 {\cal E}_0 + 3 \gamma \hat m ~ \, + \, \Pi_N
\en
where $\Pi_N$ is the self-energy operator encoding the nucleon mass shift 
due to the meson cloud. Two diagrams contribute to $\Pi_N$ at one loop: 
the meson cloud (Fig.2a) and the meson exchange diagram (Fig.2b): 
\eq \label{selfen}
\Pi_N \, = \,  
\!\! \sum_{\Phi=\pi,K,\eta} \!\!\! d_N^{\Phi \, (1)} \, \Pi_{\Phi}^{(1)}~
 +    \! \sum_{\Phi=\pi,K,\eta} \!\!\! d_N^{\Phi \, (2)} \, 
\Pi_{\Phi}^{(2)}
\end{eqnarray}
where $\Pi_{\Phi}^{(1)}$ and $\Pi_{\Phi}^{(2)}$ are the self-energies  
corresponding to the diagrams of Fig.2a and Fig.2b, respectively. 
In accordance with the FH theorem (\ref{FHTh}), the vertex functions 
$\Gamma_{\Phi}^{(i)}$ are related  to the partial derivative of the  
self-energies $\Pi_{\Phi}^{(i)}$ with respect to $\hat{m}$: 
\eq
\Gamma_{\Phi}^{(i)} \, = \, \hat{m} \, 
\frac{\partial}{\partial\hat{m}} \, \Pi_{\Phi}^{(i)} \, . 
\en
The recoupling coefficients $d_N^{\Phi \, (i)}$ of Eq. (\ref{selfen}), 
defining the partial contributions of the $\pi$, $K$, and $\eta$-meson 
cloud to the energy shift of the nucleon, are therefore the same as in 
Eq.~(\ref{sigma_piN_sea}). 

We also consider quantities which partially incorporate
information about the strangeness content of the nucleon: 
the strangeness $y_N$, the kaon-nucleon sigma-terms 
$\sigma_{KN}^{u}  \equiv \sigma_{KN}^{(1)}$, 
$\sigma_{KN}^{d}$, $\sigma_{KN}^{(2)}$, $\sigma_{KN}^{I=0}$ and 
$\sigma_{KN}^{I=1}$, the eta-nucleon sigma-term $\sigma_{\eta N}$ 
and the respective scalar form factors 
$\sigma_{KN}^{u}(Q^2) \equiv \sigma_{KN}^{(1)}(Q^2)$, 
$\sigma_{KN}^{d}(Q^2)$, $\sigma_{KN}^{(2)}(Q^2)$, 
$\sigma_{KN}^{I=0}(Q^2)$, $\sigma_{KN}^{I=1}(Q^2)$ and 
$\sigma_{\eta N}(Q^2)$ (in particular their slopes). These quantities 
of interest are defined by the standard formulas:
\eq\label{sigma_all}
& &\sigma_{KN}^{u} \equiv \sigma_{KN}^{(1)} =  
\frac{\hat m + m_s}{2} \langle p| S_{u + s}^{PCQM} |p \rangle \,, 
\nonumber\\
& &\sigma_{KN}^{d} = 
\frac{\hat m + m_s}{2} \langle p| S_{d + s}^{PCQM} |p \rangle \,, 
\nonumber\\
& &\sigma_{KN}^{(2)} \equiv 2 \sigma_{KN}^{(d)} - \sigma_{KN}^{(u)}  =  
\frac{\hat m + m_s}{2} \langle p| 2 \, S_{d + s}^{PCQM} - S_{u + s}^{PCQM} 
|p\rangle \,, \nonumber\\
& &\sigma_{KN}^{I=0} = \frac{\sigma_{KN}^u + \sigma_{KN}^d}{2} = 
\frac{\hat m + m_s}{4} \langle p| S_{u + s}^{PCQM} \, 
+ \, S_{d + s}^{PCQM}|p \rangle \,,
 \nonumber\\
& & \sigma_{KN}^{I=1} = \frac{\sigma_{KN}^u - \sigma_{KN}^d}{2} = 
\frac{\hat m + m_s}{4}  \langle p| S_{u - d}^{PCQM} |p \rangle \,, \\
& &\sigma_{\eta N} = 
\frac{1}{3} \langle p|\hat m \, S_{u + d}^{PCQM} \, + \, 4 \, m_s \, 
S_s^{PCQM}  |p \rangle \,, \nonumber\\
& &F_S = \frac{1}{2} \, \langle p| S_{u - s}^{PCQM} |p \rangle \,, 
\nonumber\\
& &D_S = \frac{1}{2} \, \langle p| S_{u - d}^{PCQM} \, - \, 
S_{d - s}^{PCQM} |p \rangle \,, \nonumber\\ 
& &y_N = 2 \, \frac{ \langle p| S_s^{PCQM} |p \rangle}
{\langle p| S_{u + d}^{PCQM} |p \rangle }\,, 
\nonumber
\en
where $|p \rangle$ denotes a one-proton state normalized to unity 
$\langle p|p \rangle = 1$. The generic matrix element 
$\langle p| S_q^{PCQM} |p \rangle$ is again set up by the expression
\eq
\langle p| S_q^{PCQM} |p \rangle  &\doteq&         
 {}^p \! \langle \phi_0| \sum_{n=0}^{2} \frac{i^n}{n!}\! 
 \int \, d^3x \, d^4x_1 \, \ldots \, d^4x_n \, \, 
T[{\cal L}_I(x_1) \ldots {\cal L}_I(x_n) \, S_q^{PCQM}(\vec{x}) \, ]
 | \phi_0 \rangle^p_c \, . 
\en  
The scalar nucleon form factors and their respective slopes are defined 
in analogy with Eqs.~(\ref{piN_pert}) and (\ref{piN_slope}), where latter 
equation will be set up in the forthcoming section.  

\section{Results}
We first discuss the results for the scalar nucleon form factor 
$\sigma_{\pi N}(Q^2)$ and the corresponding $\pi N$ sigma-term. 
In Ref.~\cite{Lyubovitskij:2000sf} we calculated the $\sigma_{\pi N}$ 
sigma-term with the quark propagator restricted to the ground state.
The result we obtained there was $45 \pm 5$ MeV, where the variation of 
the value is due to a change of the range parameter $R$. For the central 
value of $R=0.6$ fm we obtain $\sigma_{\pi N} = 42.5$ MeV, where the 
valence-quark contribution is $\sigma_{\pi N}^{val} = 13.1$ MeV 
(i.e. 1/3 of the total value) and the meson-cloud contribution is 
dominated by the pions with $\sigma_{\pi N}^{sea} = 29.4$ MeV 
(i.e. 2/3 of total value). 

Inclusion of the excited quark states obviously does not change the 
result for $\sigma_{\pi N}^{val}$ but leads to an increase for
$\sigma_{\pi N}^{sea}$ from 29.4 MeV to 41.6 MeV. Hence, we obtain a 
sizable increase of the sigma term by about 12 MeV leading to the final
value of $\sigma_{\pi N}$ =   54.7 MeV. Again, the main contribution is 
due to pion loops, whereas kaon and eta loops are strongly suppressed. 
The obtained value of 54.7 MeV is still comparable to the upper limit of 
the canonical result $45 \pm 8$ MeV obtained in 
Ref.~\cite{Gasser:1990ce}. It is also in agreement with the result 
obtained in the framework of relativistic baryon ChPT up to 
next-next-to-leading order (NNLO) based on an extrapolation of this 
observable from two-flavor lattice QCD results: 
$\sigma_{\pi N} = 53 \pm 8$ MeV at the physical value of the pion 
mass~\cite{Procura:2003ig}. For a more detailed comparison to other 
theoretical approaches we relegate to the recent 
paper~\cite{Schweitzer:2003sb}. 

Our final results are compiled as: 

\eq\label{sigma_piN_fin}
& &\sigma_{\pi N} = 54.7 \,\,\, {\rm MeV}, \hspace*{.5cm} 
\sigma_{\pi N}^{val} = 13.1 \,\,\, {\rm MeV}, \hspace*{.5cm} 
\sigma_{\pi N}^{sea} = 41.6 \,\,\, {\rm MeV}, \\
& &\sigma_{\pi N}^{\pi} = 39.4 \,\,\, {\rm MeV}, \hspace*{.5cm} 
\sigma_{\pi N}^{K} = 2.1 \,\,\, {\rm MeV}, \hspace*{.8cm} 
\sigma_{\pi N}^{\eta} = 0.1 \,\,\, {\rm MeV} \,  ,\nonumber
\en 
where the explicit meson loop contributions are made explicit in the 
last line.

In Fig.3 we plot the behavior of the scalar nucleon form factor 
$\sigma_{\pi N}(Q^2)$ in the space-like region up to 0.5 GeV$^2$.  
The partial contributions of the valence quarks and the meson cloud are 
indicated separately. The meson cloud dominates the scalar nucleon form 
factor for small $Q^2 < 0.15$ GeV$^2$, while the valence quarks contribute 
mostly at large $Q^2 > 0.3$ GeV$^2$. 
The limiting value for 
$\sigma_{\pi N}(Q^2)$ at $Q^2=0$ is nothing but the $\pi N$ sigma-term 
as discussed above. The slope of the scalar nucleon form factor is 
defined by the standard formula: 
\eq\label{piN_slope}
\langle r^2 \rangle^S_N \, \doteq 
\langle r^2 \rangle^S_{\pi N} \, =  \, - \, 
\frac{6}{\sigma_{\pi N}(0)} \frac{d\sigma_{\pi N}(Q^2)}{dQ^2}
\Bigg|_{\displaystyle{Q^2 = 0}} \, . 
\en
Our result is 
\eq
\langle r^2 \rangle^S_N = 1.5 \,\, {\rm fm}^2 \, 
\en
which is comparable to the model-independent prediction of 
Ref.~\cite{Gasser:1990ce}:  $\langle r^2 \rangle^S_N \simeq 1.6$ fm$^2$. 
As pointed out in Ref.~\cite{Gasser:1990ce}, the result for the range
of the scalar form factor is larger 
than the charge radius of nucleon associated with the isovector form 
factor since "the scalar current $\bar u u + \bar d d$ is much more 
sensitive to the pion halo surrounding the nucleon than the 
electromagnetic current".  
In Fig.4 we compare our prediction for the scalar form factor 
$\sigma_{\pi N}(Q^2)$ to results of other theoretical calculations:  
lattice QCD~\cite{Dong:1995ec}, soliton models  
(NJL-soliton model~\cite{Kim:1995hu} and chiral quark soliton  
model~\cite{Schweitzer:2003sb}) and the relativistic baryon chiral  
perturbation theory~\cite{Fuchs:2003kq}. The lattice result for  
$\sigma_{\pi N}(Q^2)$~\cite{Dong:1995ec} with  
$\sigma_{\pi N}(0) = 49.7 \pm 2.6$ MeV decreases slower than ours.    
Their prediction for the slope  
$\sqrt{\langle r^2 \rangle^S_N} = 0.85(4)$ fm is also smaller than our  
estimate and the model-independent results of Ref.~\cite{Gasser:1990ce}.  
The form factor $\sigma_{\pi N}(Q^2)$ calculated  
in Ref.~\cite{Kim:1995hu} starts from 40.8 MeV at zero recoil, crosses  
our form factor at around 0.17 GeV$^2$ and decreases slower than our  
curve. The result for $\langle r^2 \rangle^S_N = $ 1.5 fm$^2$ coincides  
with ours. The result of Ref.~\cite{Schweitzer:2003sb} has the same  
shape but is shifted in average by about $\sim 15$ MeV upward with  
respect to our result. The corresponding values of $\sigma_{\pi N}(0)$  
and $\langle r^2 \rangle^S_N $ are 67.9 MeV and 1 fm$^2$.   
The scalar form factor calculated in Ref.~\cite{Fuchs:2003kq} is smaller  
than ours in magnitude in the Euclidean region and decreases very  
quickly. It vanishes at 0.2 GeV.  
The value of $\sigma_{\pi N}(0) = 40.5$ MeV calculated by neglecting  
higher-order corrections is close to the canonical  
value of $\sigma_{\pi N}(0) = 45 \pm 8$ MeV~\cite{Gasser:1990ce}. 

Next we extrapolate the scalar nucleon form factor to the 
time-like region $t = - Q^2$ for small $t$ by using the linear 
approximation: 
\eq\label{lin_ext}
\sigma_{\pi N}(t) \, = \, \sigma_{\pi N}(0) \, 
\biggl( 1 \, + \, \frac{1}{6} \, \langle r^2 \rangle^S_N \, \cdot \, t 
\, + \, O(t^2) \, \biggr) \, . 
\en
With Eqs.~(\ref{lin_ext}) and (\ref{piN_slope}) 
we obtain for the difference 
\eq
\Delta_\sigma = \sigma_{\pi N}(2M_{\pi}^2) 
- \sigma_{\pi N}(0) = 13.8 \,\,\, \mbox{MeV} 
\en
which is comparable to the canonical value of 
$\Delta_\sigma = 15.2 \pm 0.4$ MeV deduced by dispersion-relation 
techniques~\cite{Gasser:1990ce}  and to the results obtained in ChPT: 
$\Delta_\sigma = 14.0$ MeV $+ 2 M^4 \bar e_2$~\cite{Becher:1999he} and  
$\Delta_\sigma = 16.9$ MeV $+ 2 M^4 \beta$~\cite{Fuchs:2003kq}. 
The value of the $\sigma_{\pi N}(t)$ at the CD point 
\eq
\sigma_{\pi N}(2M_{\pi}^2) = 68.5 \,\,\, \mbox{MeV}
\en 
is comparable to the upper limit ($68$ MeV) of the value deduced in 
Ref.~\cite{Gasser:1990ce}, close to the central value ($64$ MeV) 
extracted from the analysis of pion-nucleon scattering 
data~\cite{Koch:pu} and smaller than the central values of the recent 
analyses~\cite{Kaufmann:dd,Olsson:1999jt,Pavan:2001wz}. 
Our result for $\Delta_\sigma$ is close to the prediction of soliton 
models 14.7 MeV~\cite{Schweitzer:2003sb} and 18.18 MeV~\cite{Kim:1995hu}. 
The result of lattice QCD with $\Delta_\sigma = 6.6 \pm 0.6$ MeV is 
smaller than our estimate and previous results including the canonical 
value of Ref.~\cite{Gasser:1990ce}. 

Our complete results for the static nucleon properties associated with 
the scalar density condensates of all three flavors are summarized in 
Table I. There we indicate explicitly the partial contributions of the 
valence quarks and the meson cloud: pion, kaon and eta-meson contribution.
Table I also contains the sigma-terms related to the strangeness content 
of the nucleon. Our result for the isosinglet $KN$ sigma-term 
$\sigma_{KN}^{I=0}=386.3$ MeV is in rather good agreement with other 
theoretical predictions: $389(14)$ MeV~\cite{Dong:1995ec} (lattice QCD), 
$2.83 M_\pi = 395$ MeV~\cite{Lee:1994jj} (chiral analysis of $KN$ 
scattering). The prediction for the $\sigma_{KN}^{(1)}$ sigma-term of 
$419.2$ MeV is close to the the result of the lattice-regularized 
ChPT~\cite{Borasoy:2002hz}: 300-400 MeV, where the lattice space 
parameter is varied from $\sim 1$ GeV to $10$ GeV. Our results for the 
sigma-terms $\sigma_{KN}^{(1)}$ and $\sigma_{KN}^{(2)}=287.6$ MeV are 
comparable to the prediction of heavy baryon ChPT~\cite{Borasoy:1998uu} 
(decuplet states included in the fit): $\sigma_{KN}^{(1)} = 380 \pm 40$ 
MeV and $\sigma_{KN}^{(2)} = 250 \pm 30$ MeV. Our result for the 
strangeness content of the nucleon of $y_N = 0.09$ is still smaller 
than the result of Ref.~\cite{Gasser:1990ce}: $y_N = 0.2$. 

In the following we focus on the discussion of the isovector sigma-term 
$\sigma_{KN}^{I=1}$. 
In this case the relevant scalar density operator is given by 
\begin{equation}
S_{u-d}^{\rm PCQM} \, = \, \bar u u - \bar d d \, + \, 
B \, \biggl\{ \, K^+ \, K^- \, - \, K^0 \, \bar K^0 \, \biggr\} \,, 
\label{eqn:optmupd}
\end{equation}
where the first two terms correspond to the valence quarks. 
The next two terms indicate the isovector combination of the kaon with 
$K^+K^- - K^0 \bar K^0$. The value we obtain for this sigma term is 
\eq\label{KNI1} 
\sigma_{KN}^{I=1} = 28.4 \,\, ({\rm Val}) + 4.5 \,\, 
({\rm kaon \, cloud}) = 32.9 ~ \mbox{MeV} 
\, . 
\en
Our result for the isovector kaon-nucleon sigma-term 
is smaller than the phenomenological value $\sim 50$ MeV derived in 
Ref.~\cite{Gasser:2000wv} using the baryon mass formulas: 
\eq
\sigma_{KN}^{I=1} \sim \frac{m_s + \hat{m}}{m_s - \hat{m}} \, 
\frac{m_\Xi^2 - m_\Sigma^2}{8 m_P} = 48 \,\, {\rm MeV} 
 \sim 50 \,\,  {\rm MeV}  \, .
\en
Note, the pion cloud does not contribute to $\sigma_{KN}^{I=1}$ 
(\ref{KNI1}), since a two-pion configuration with orbital 
momentum $L=0$ only occurs with the isospin combination $I=0$ or $I=2$.
Because of the absence of pion loop contributions $\sigma_{KN}^{I=1}$ 
is rather small when compared to other $KN$ sigma-terms.

In Figs.5-10 we plot the behavior of the scalar nucleon 
form factors  $\sigma_{K N}^{u}(Q^2)$, $\sigma_{K N}^{d}(Q^2)$, 
$\sigma_{K N}^{(2)}(Q^2)$, $\sigma_{K N}^{I=0}(Q^2)$, 
$\sigma_{K N}^{I=1}(Q^2)$ and 
$\sigma_{\eta N}(Q^2)$ in the space-like region up to 0.5 GeV$^2$.  
As was the case for the $\sigma_{\pi N}(Q^2)$ form factor, here we also 
indicate the partial contributions of the valence quarks and the meson 
cloud. The set of scalar nucleon form factors can be separated into three 
groups. The first group includes the form factors $\sigma_{\pi N}(Q^2)$, 
$\sigma_{K N}^{u}(Q^2)$, $\sigma_{K N}^{d}(Q^2)$, 
$\sigma_{K N}^{I=0}(Q^2)$, which is characterized by the dominance of 
the meson cloud contribution for small $Q^2 < 0.15$ GeV$^2$ and of the 
valence quarks for large $Q^2 > 0.3$ GeV$^2$. The second group includes 
the $\sigma_{K N}^{I=1}(Q^2)$ form factor which is dominated by the 
valence quark contribution in the whole $Q^2$ region. Finally, the third 
group includes the form factors $\sigma_{K N}^{(2)}(Q^2)$ and 
$\sigma_{\eta N}(Q^2)$ which are dominated by the meson cloud contribution. 

For completeness, we also determined the slopes of these form factors using: 
\eq
\langle r^2 \rangle^S_{M N} \, = \, - \, 
\frac{6}{\sigma_{M N}(0)} \frac{d\sigma_{M N}(Q^2)}{dQ^2}
\Bigg|_{\displaystyle{Q^2 = 0}} \,  , 
\en
where our results are also given in Table I. 
Unfortunately, the model cannot be applied to the calculation of the 
$\sigma_{K N}^{u}$ form factor at the CD point with $t=-Q^2 = 2 M_K^2$ 
since it is kinematically far from zero recoil. This quantity is relevant
for the ongoing DEAR experiment~\cite{Curceanu:2000dc}. 

\section{Conclusions}

We have updated our analysis of the meson-nucleon sigma-terms applying 
the perturbative chiral quark model which is based on an effective chiral 
Lagrangian. As a new feature we include excited states 
in the quark propagator, which, in the context of loop diagrams, can
play an important role to enhance the considered quantities.
We presented a comprehensive calculation for various flavor combinations
of the scalar nucleon form factors and the respective low-energy
properties such as slopes and sigma terms.

In particular we showed that inclusion of excited states leads to an
increase of the $\pi N$ sigma term from $\sim 45$~MeV to $\sim 55$~MeV. 
The main contribution to this quantity arises from the pion cloud. We
also determined the 
scalar nucleon form factor $\sigma_{\pi N}(Q^2)$ in the space-like 
region up to 0.5 GeV$^2$ and extrapolates it to the Cheng-Dashen point 
$t = - Q^2 = 2 M_\pi^2$. Our result for the difference 
 $\Delta_\sigma = \sigma_{\pi N}(2M_{\pi}^2) - \sigma_{\pi N}(0) 
= 13.8$ MeV is in good agreement with the canonical value 
$15.2 \pm 0.4$ MeV~\cite{Gasser:1990ce} 
and the results obtained in ChPT~\cite{Becher:1999he,Fuchs:2003kq}.
 
We also presented a detailed analysis of quantities related to the
strangeness content of the nucleon such as $KN$ and $\eta N$ sigma terms, 
$y_N$, $KN$ and $\eta N$ scalar form factors and their respective slopes.
To gain a deeper understanding of the various meson-nucleon sigma terms, 
further efforts are needed both from theoretical and experimental side.

\vspace*{.5cm}

{\bf Acknowledgments}

\vspace*{.5cm}

\noindent
This work was supported by the Deutsche Forschungsgemeinschaft (DFG) 
under contracts FA67/25-3 and GRK683 and by the State of 
Baden-W\"{u}rttemberg, LFSP "Quasiparticles". 

\appendix 
\section{Aspects of chiral Lagrangian}

Using the standard phase conventions for the pseudoscalar meson fields 
$\Phi_P$ and matrices $\lambda_P$ are given by~\cite{Gasser:1982ap}  
\begin{eqnarray}
&&\Phi_{\pi^\pm} = - \frac{1}{\sqrt{2}} (\Phi_1 \mp i \Phi_2)\,, 
\hspace*{.5cm} 
\Phi_{K^\pm} = - \frac{1}{\sqrt{2}} (\Phi_4 \mp i \Phi_5)\,,\nonumber\\
&&\Phi_{K^0} = - \frac{1}{\sqrt{2}} (\Phi_6 - i \Phi_7)\,, 
\hspace*{.5cm} 
\Phi_{\bar K^0} = - \frac{1}{\sqrt{2}} (\Phi_6 + i \Phi_7)\,,\nonumber\\ 
&&\Phi_{\pi^0} = \Phi_3 \cos\varepsilon + \Phi_8 \sin\varepsilon\,, 
\hspace*{.5cm}   
\Phi_{\eta} = - \Phi_3 \sin\varepsilon + \Phi_8 \cos\varepsilon\,, 
\nonumber\\ 
&&\\
&&\lambda_{\pi^\pm} = - \frac{1}{\sqrt{2}} (\lambda_1 \pm i \lambda_2)\,, 
\hspace*{.5cm} 
\lambda_{K^\pm} = - \frac{1}{\sqrt{2}} (\lambda_4 \pm i \lambda_5)\,,
\nonumber\\
&&\lambda_{K^0} = - \frac{1}{\sqrt{2}} (\lambda_6 + i \lambda_7)\,, 
\hspace*{.5cm} 
\lambda_{\bar K^0} = - \frac{1}{\sqrt{2}} (\lambda_6 - i \lambda_7)\,,
\nonumber\\ 
&&\lambda_{\pi^0} = \lambda_3 \cos\varepsilon + \lambda_8 \sin\varepsilon\,, 
\hspace*{.5cm}   
\lambda_{\eta} = - \lambda_3 \sin\varepsilon + \lambda_8 \cos\varepsilon\,. 
\nonumber 
\end{eqnarray}
The $\pi^0$-$\eta$ mixing angle $\varepsilon$ with 
\begin{eqnarray}
\tan 2\varepsilon \, = \, \frac{\sqrt{3}}{2} \, 
\frac{m_d - m_u}{m_s - \hat{m}}\,, 
\hspace*{.5cm} \hat{m} \, = \, \frac{1}{2} \, (m_u \, + \, m_d)    
\end{eqnarray} 
is fixed from the diagonalization of the pseudoscalar meson mass term 
\begin{eqnarray}
\frac{B}{2} \, {\rm Tr}[\hat \Phi^2  {\cal M} ] \, = \, 
\frac{1}{2} \, \sum\limits_P \, M_P^2 \, \Phi^2_P \,.  
\end{eqnarray}
The masses of the mesons at leading order are: 
\begin{eqnarray}\label{Meson_Masses}
&&M_{\pi\pm}^2 = 2 \hat{m} B\,, \hspace*{.5cm} 
M_{K\pm}^2 = (m_u + m_s) B\,, \hspace*{.5cm} 
M_{K^0}^2 = M_{\bar{K}^0}^2 = (m_d + m_s) B\,,  \\
&&M_{\pi^0}^2 = 2 \hat{m} B - \frac{4}{3} (m_s - \hat{m}) 
B \, \frac{\sin^2\varepsilon}{\cos 2\varepsilon}\,, \hspace*{.5cm} 
M_{\eta}^2 = \frac{2}{3}(\hat{m} + 2 m_s) B + \frac{4}{3} (m_s - \hat{m}) 
B \,\frac{\sin^2\varepsilon}{\cos 2\varepsilon}\,.\nonumber
\end{eqnarray}
Finally, with the use of the set $\{\Phi_P, \lambda_P\}$ the linearized 
Lagrangian (\ref{linearized_L}) can be written as: 
\begin{eqnarray}\label{linearized_LP}
{\cal L}_{\rm eff}(x) &=&
\bar{\psi}(x) \, [ \, i \not\!\partial - S(r) - \gamma^0 V(r) 
- {\cal M} \, ] \, \psi(x) 
- \frac{1}{2} \sum\limits_{P} \Phi_P(x) [\Box + M_P^2] \Phi_P(x)  
\nonumber\\
&-& \bar{\psi}(x) \, i \gamma^5 \, \frac{S(r)}{F} \, 
\sum\limits_{P} \, \Phi_P(x) \, \lambda_P \, \psi(x) 
\end{eqnarray}
where $\Box = \partial_\mu \partial^\mu$.

\newpage 


\begin{table}[t]
\caption{Meson-nucleon sigma terms and related quantities}  
\begin{center}
\def\arraystretch{1.5}
\begin{tabular}{cccccc}
Quantity & Val & $\pi$ & $K$ & $\eta$ & Total  \\
 \hline 
$\sigma_{\pi N}$ \,\,\,({\rm MeV}) 
& 13.1 & 39.4 & 2.1 & 0.1 & 54.7 \\
\hline
$\Delta_\sigma $ \,\,\,({\rm MeV}) 
& 1 & 12.5 & 0.3 & 0.02 & 13.8 \\
\hline
$\sigma_{K N}^u$ \,\,\,({\rm MeV}) 
& 113.8 & 256.2 & 44.7 & 4.5 & 419.2 \\
\hline
$\sigma_{K N}^d$ \,\,\,({\rm MeV}) 
& 56.9 & 256.2 & 35.8 & 4.5 & 353.4 \\
\hline
$\sigma_{K N}^{(2)}$ \,\,\,({\rm MeV}) 
& 0 & 256.2 & 26.8 & 4.5 & 287.5 \\
\hline
$\sigma_{K N}^{I=0}$ \,\,\,({\rm MeV}) 
& 85.4 & 256.2 & 40.2 & 4.5 & 386.3 \\
\hline
$\sigma_{K N}^{I=1}$ \,\,\,({\rm MeV}) 
& 28.4 & 0 & 4.5 & 0 & 32.9\\
\hline
$\sigma_{\eta N}$ \,\,\,({\rm MeV}) 
& 4.4 & 13.1 & 69.4 & 9.4 & 96.3 \\
\hline
$\langle p|S^{PQCM}_u|p \rangle$
& 1.3 & 2.8 & 0.2 & 0.01 & 4.3 \\
\hline
$\langle p|S^{PQCM}_d|p \rangle$
& 0.6 & 2.8 & 0.1 & 0.01 & 3.5 \\
\hline
$\langle p|S^{PQCM}_s|p \rangle$
& 0 & 0 & 0.3 & 0.04 & 0.34 \\
\hline
$F_S$
& 0.63 & 1.41 & $-$ 0.05 & $-$ 0.01 & 1.97 \\
\hline
$D_S$
& 0 & $-$ 1.41 & 0.15 & 0.01 & $-$ 1.25 \\
\hline
$\langle r^2 \rangle^S_{\pi N}$ \,\,\,({\rm fm}$^2$) 
& 0.10 & 1.36 & 0.04 & 0.002 & 1.50 \\
\hline
$\langle r^2 \rangle^{S; \, (u)}_{K N}$ \,\,\,({\rm fm}$^2$) 
& 0.12 & 1.16 & 0.10 & 0.01 & 1.39 \\
\hline
$\langle r^2 \rangle^{S; \, (d)}_{K N}$ \,\,\,({\rm fm}$^2$) 
& 0.07 & 1.37 & 0.10 & 0.01 & 1.55 \\
\hline
$\langle r^2 \rangle^{S; \, (2)}_{K N}$ \,\,\,({\rm fm}$^2$) 
& 0 & 1.69 & 0.09 & 0.02 & 1.80 \\
\hline
$\langle r^2 \rangle^{S; \, I=0}_{K N}$  \,\,\,({\rm fm}$^2$) 
& 0.10 & 1.25 & 0.10 & 0.01 & 1.46 \\
\hline
$\langle r^2 \rangle^{S; \, I=1}_{K N}$ \,\,\,({\rm fm}$^2$) 
& 0.37 & 0 & 0.13 & 0 & 0.50 \\
\hline
$\langle r^2 \rangle^S_{\eta N}$  \,\,\,({\rm fm}$^2$) 
& 0.02 & 0.26 & 0.71 & 0.10 & 1.09 \\
\end{tabular}
\end{center}
\end{table}

\newpage
\begin{figure}[t]
\noindent Fig.1: Diagrams contributing to the meson-baryon sigma-terms: 

\noindent tree diagram (1a), meson cloud diagram (1b) and  
meson exchange diagram (1c).
 
\noindent Insertion of the scalar density operator is depicted 
by the symbol $" \bf \vee "$. 

\vspace*{.75cm}

\noindent Fig.2: Diagrams contributing to the baryon energy shift:  

\noindent meson cloud (2a) and meson exchange diagram (2b). 

\vspace*{.75cm}
\noindent Fig.3: Scalar nucleon form factor $\sigma_{\pi N}(Q^2)$: 
valence quark, meson cloud and total contributions.

\vspace*{.75cm}
\noindent Fig.4: Scalar nucleon form factor $\sigma_{\pi N}(Q^2)$ 
in comparison to other theoretical approaches (our result - solid line, 
chiral quark soliton model~\cite{Schweitzer:2003sb}) - dashed line, 
relativistic baryon chiral perturbation theory~\cite{Fuchs:2003kq} - 
short-dashed line, NJL-soliton model~\cite{Kim:1995hu} - dotted line and  
lattice QCD~\cite{Dong:1995ec} - dash-dotted line). 

\vspace*{.75cm}
\noindent Fig.5: Scalar nucleon form factor $\sigma_{K N}^{u}(Q^2)$: 
valence quark, meson cloud and total contributions.

\vspace*{.75cm}
\noindent Fig.6: Scalar nucleon form factor $\sigma_{K N}^{d}(Q^2)$: 
valence quark, meson cloud and total contributions.

\vspace*{.75cm}
\noindent Fig.7: Scalar nucleon form factor $\sigma_{K N}^{(2)}(Q^2)$: 
total (meson cloud) contribution. 

\vspace*{.75cm}
\noindent Fig.8: Scalar nucleon form factor $\sigma_{K N}^{I=0}(Q^2)$: 
valence quark, meson cloud and total contributions.

\vspace*{.75cm}
\noindent Fig.9: Scalar nucleon form factor $\sigma_{K N}^{I=1}(Q^2)$: 
valence quark, meson cloud and total contributions.

\vspace*{.75cm}
\noindent Fig.10: Scalar nucleon form factor $\sigma_{\eta N}(Q^2)$: 
valence quark, meson cloud and total contributions.
\end{figure}

\begin{figure}
\centering{\
\epsfig{figure=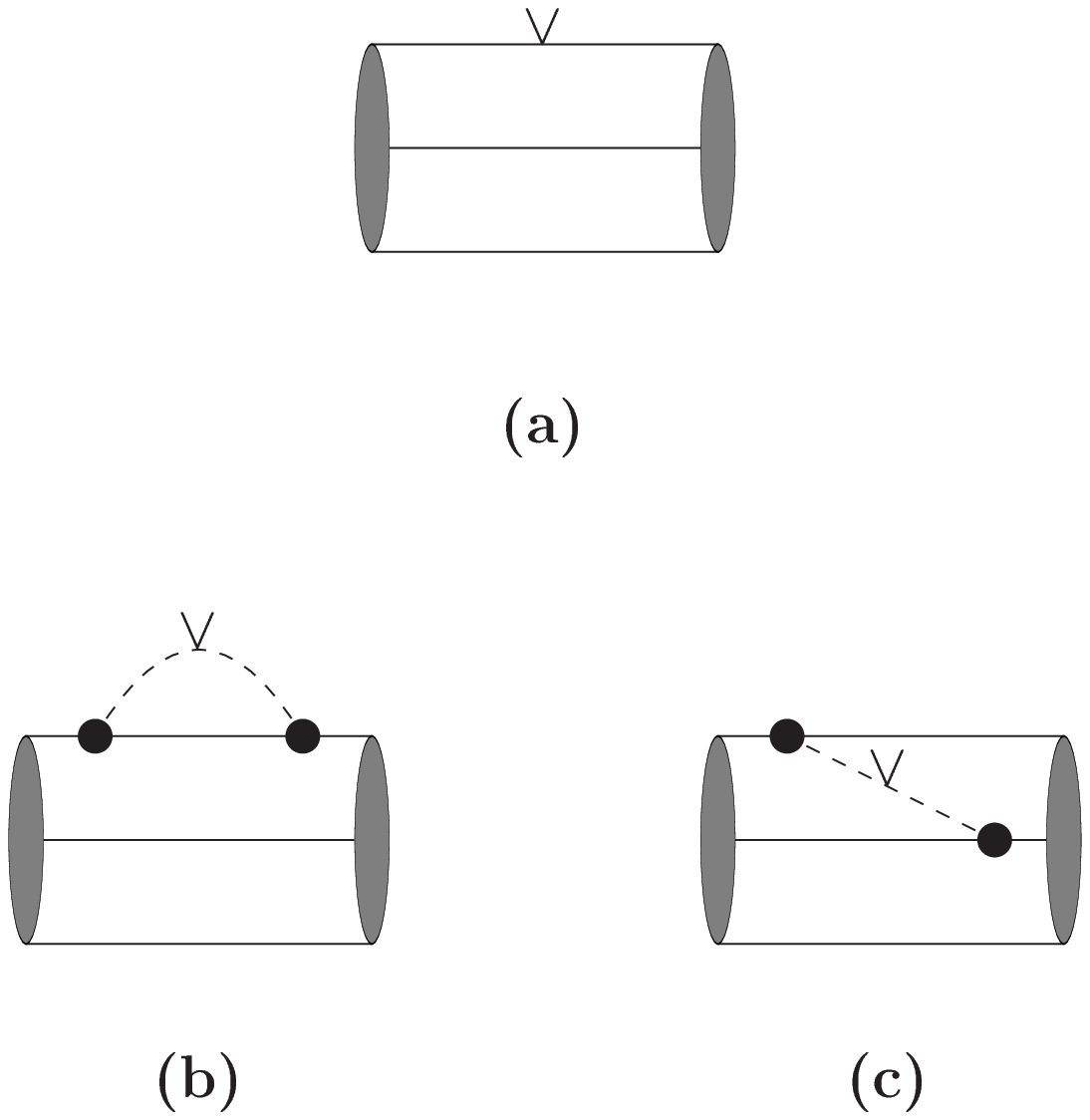,height=21cm}}

\vspace*{-8cm}

\centerline{\bf Fig.1}
\end{figure}

\begin{figure}
\centering{\
\epsfig{figure=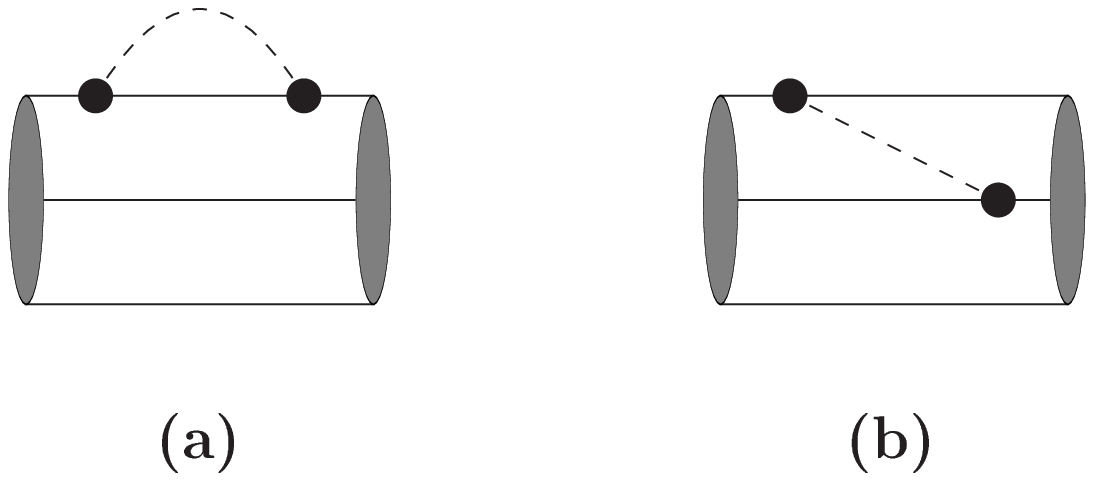,height=21cm}}
\end{figure}

\vspace*{-8cm}

\centerline{\bf Fig.2}

\newpage

\begin{figure}
\centering{\
\epsfig{figure=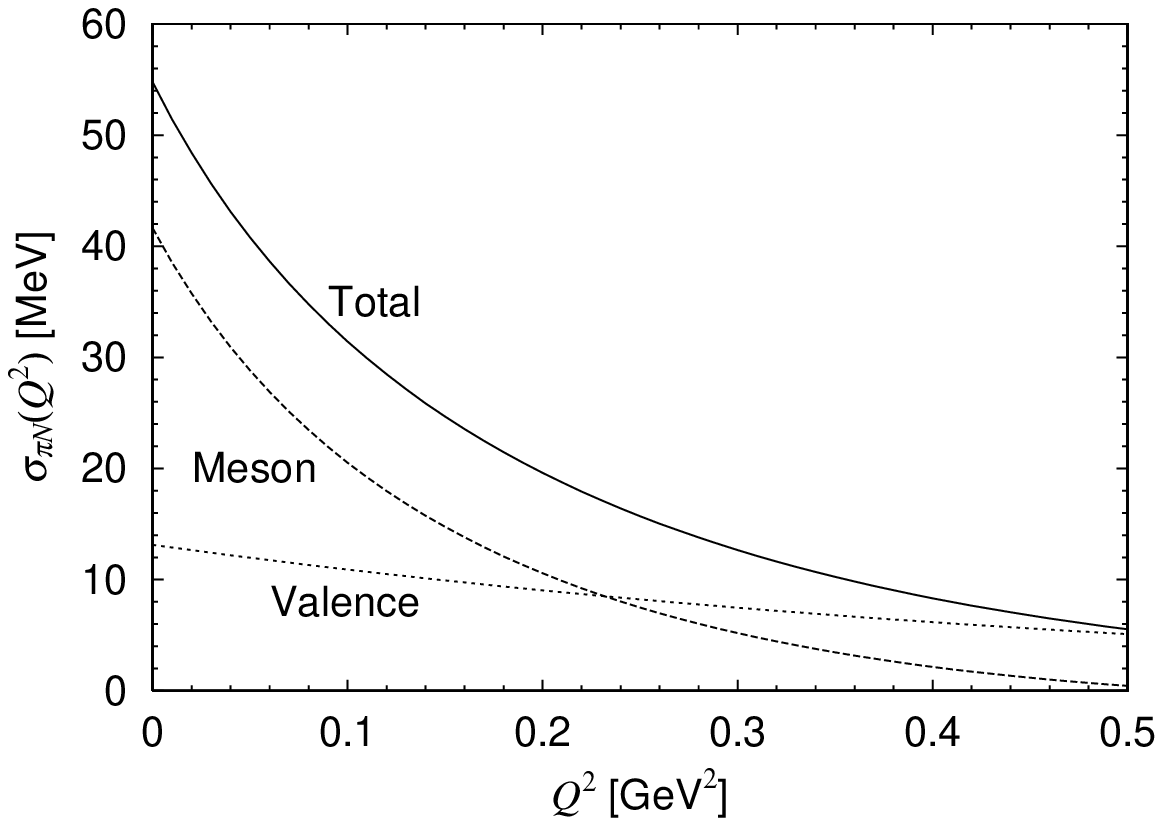,height=10cm}}
\end{figure}

\vspace*{2cm}

\centerline{\bf Fig.3}

\newpage

\begin{figure}
\centering{\
\epsfig{figure=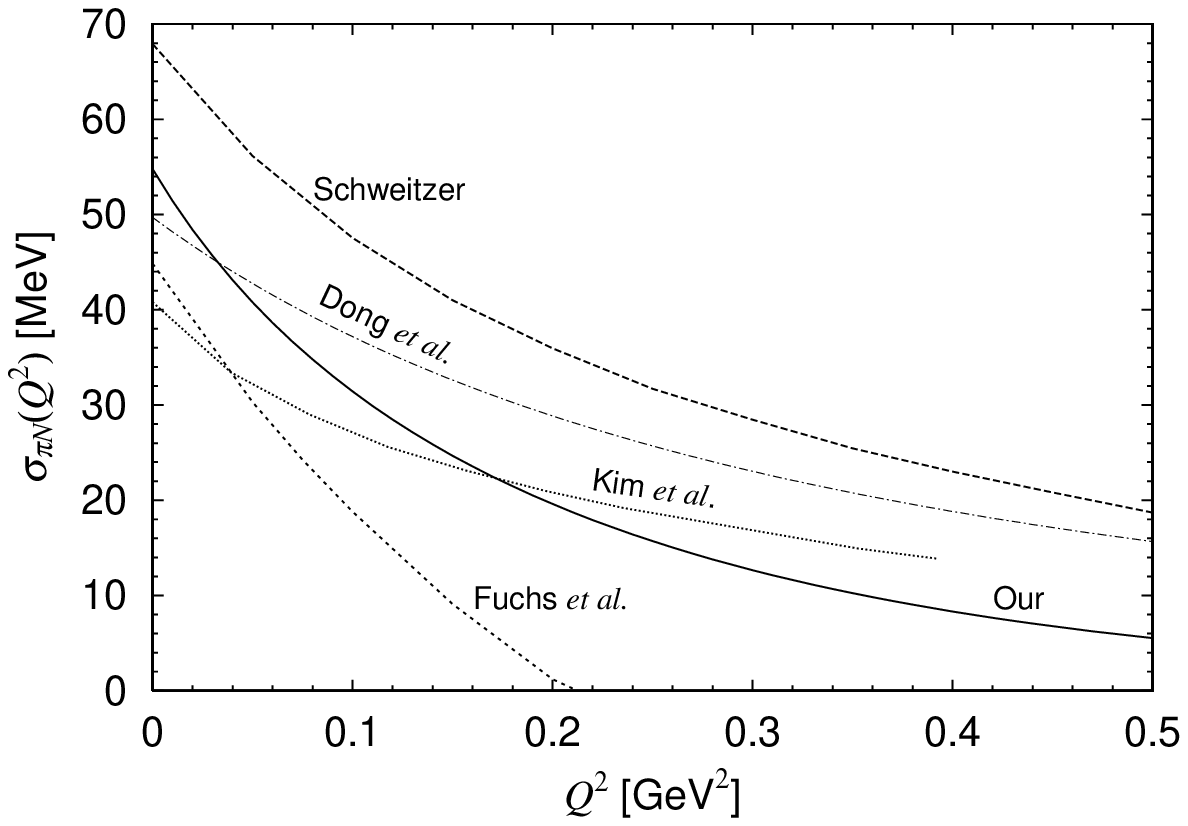,height=10cm}}
\end{figure}

\vspace*{2cm}

\centerline{\bf Fig.4}

\newpage

\begin{figure}
\centering{\
\epsfig{figure=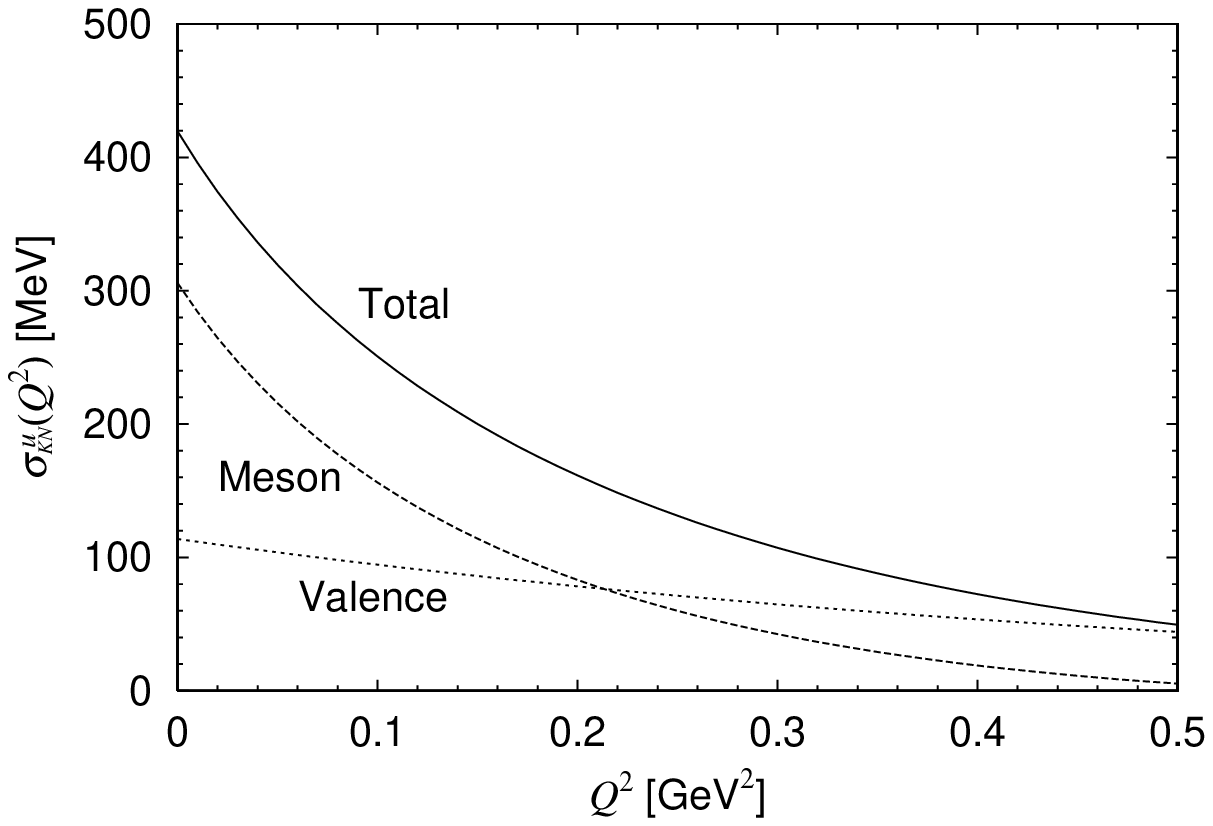,height=10cm}}
\end{figure}

\vspace*{2cm}

\centerline{\bf Fig.5}

\newpage

\begin{figure}
\centering{\
\epsfig{figure=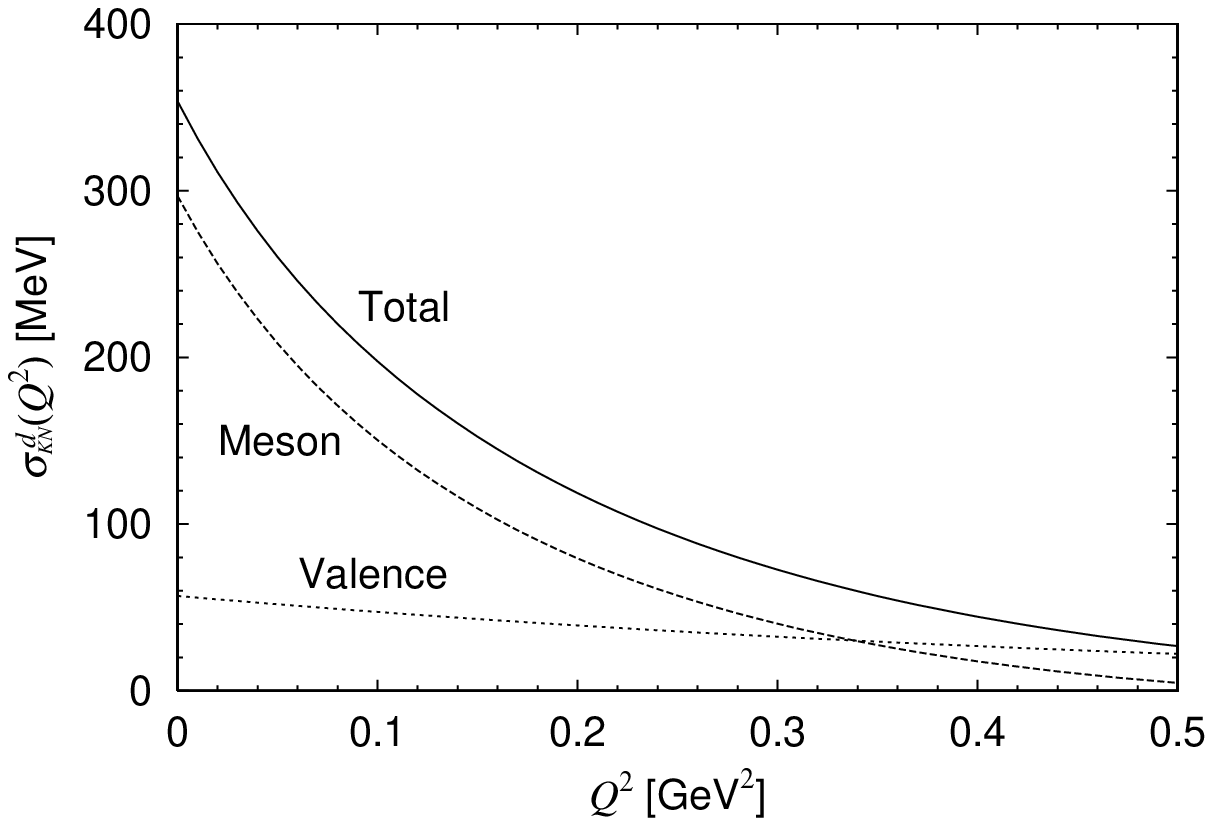,height=10cm}}
\end{figure}

\vspace*{2cm}

\centerline{\bf Fig.6}

\newpage

\begin{figure}
\centering{\
\epsfig{figure=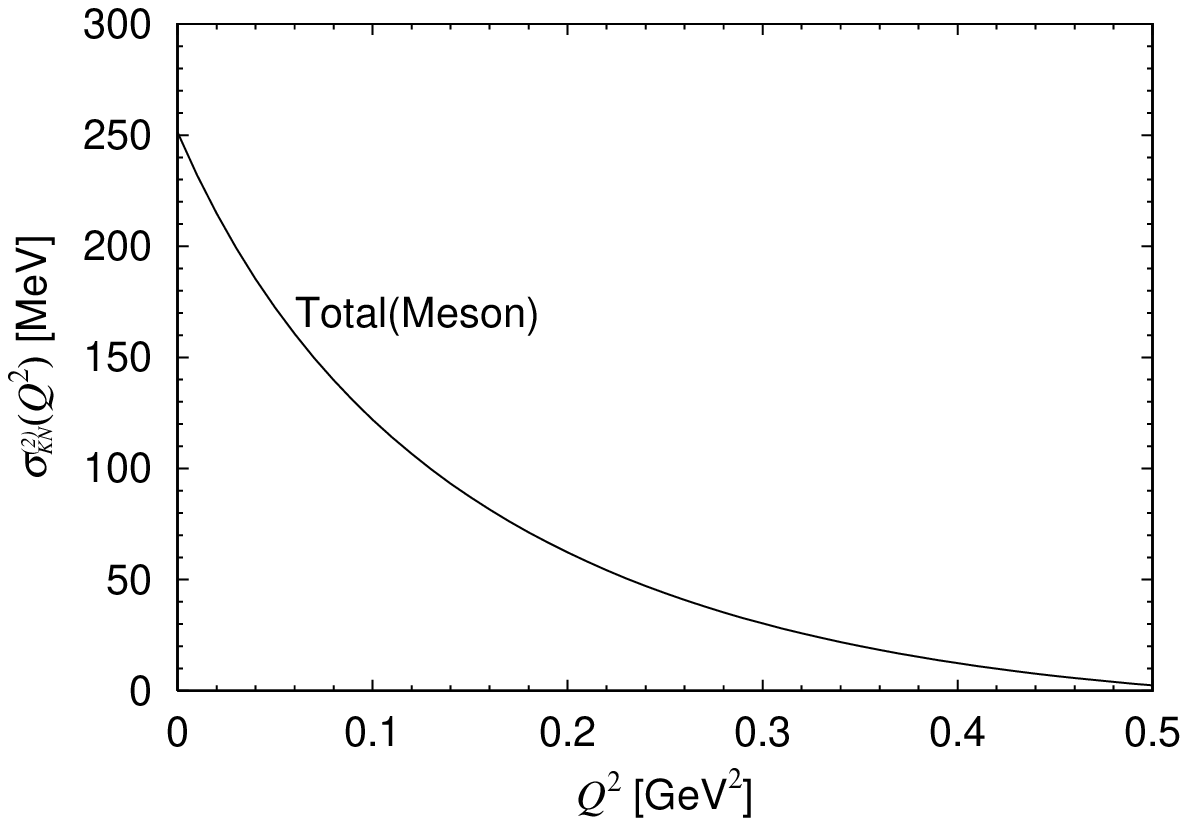,height=10cm}}
\end{figure}

\vspace*{2cm}

\centerline{\bf Fig.7}

\newpage

\begin{figure}
\centering{\
\epsfig{figure=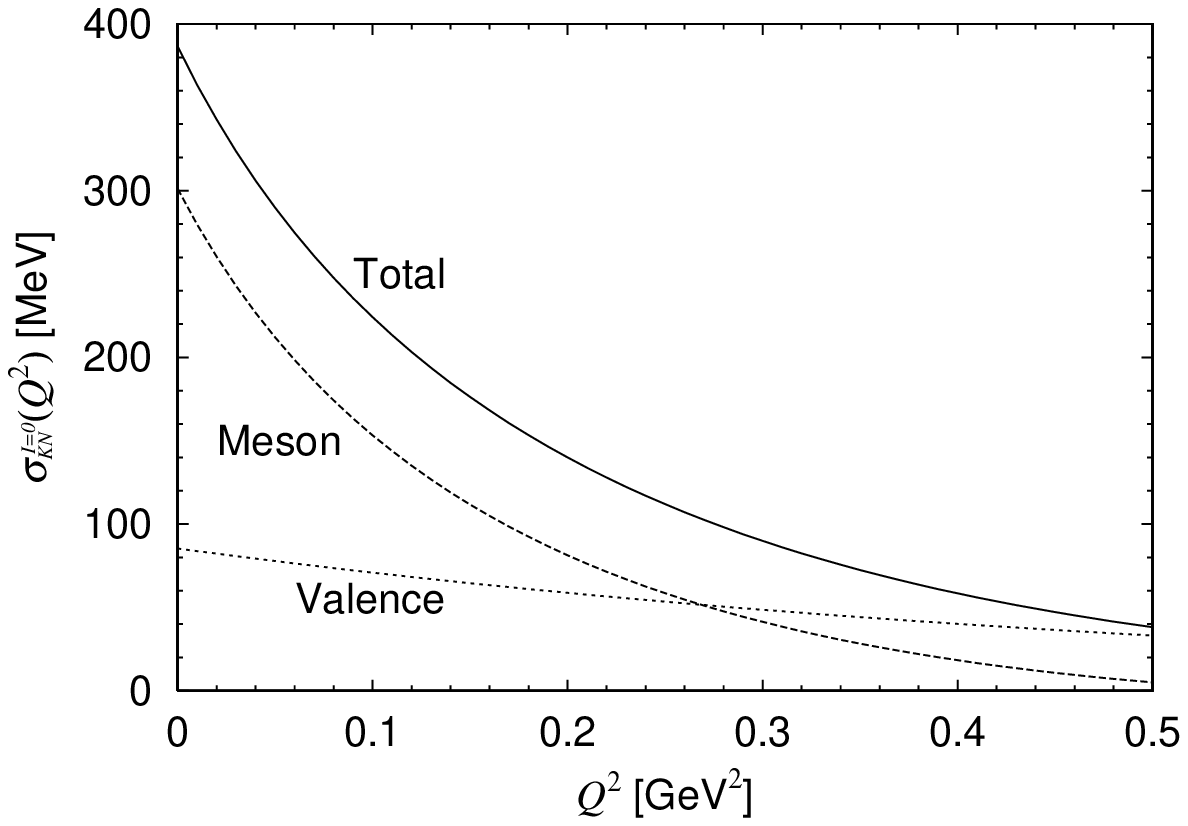,height=10cm}}
\end{figure}

\vspace*{2cm}

\centerline{\bf Fig.8}

\newpage

\begin{figure}
\centering{\
\epsfig{figure=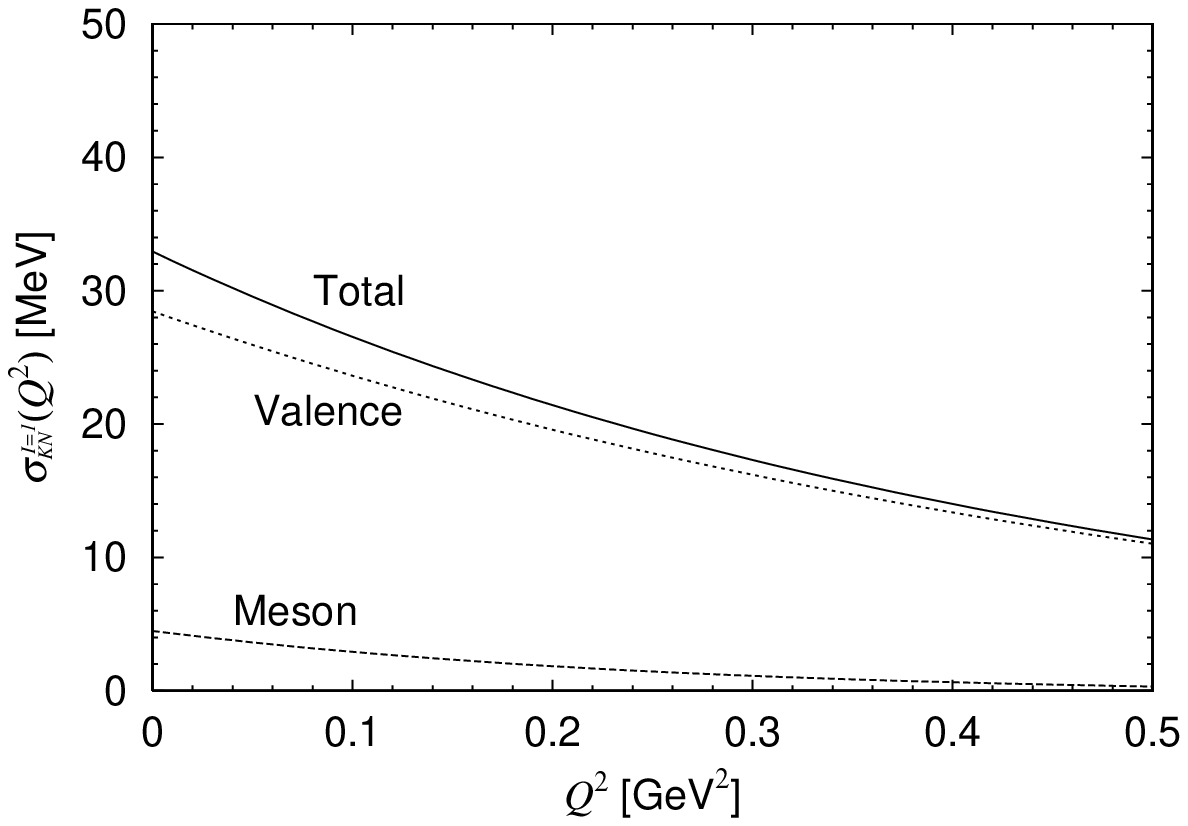,height=10cm}}
\end{figure}

\vspace*{2cm}

\centerline{\bf Fig.9}

\newpage

\begin{figure}
\centering{\
\epsfig{figure=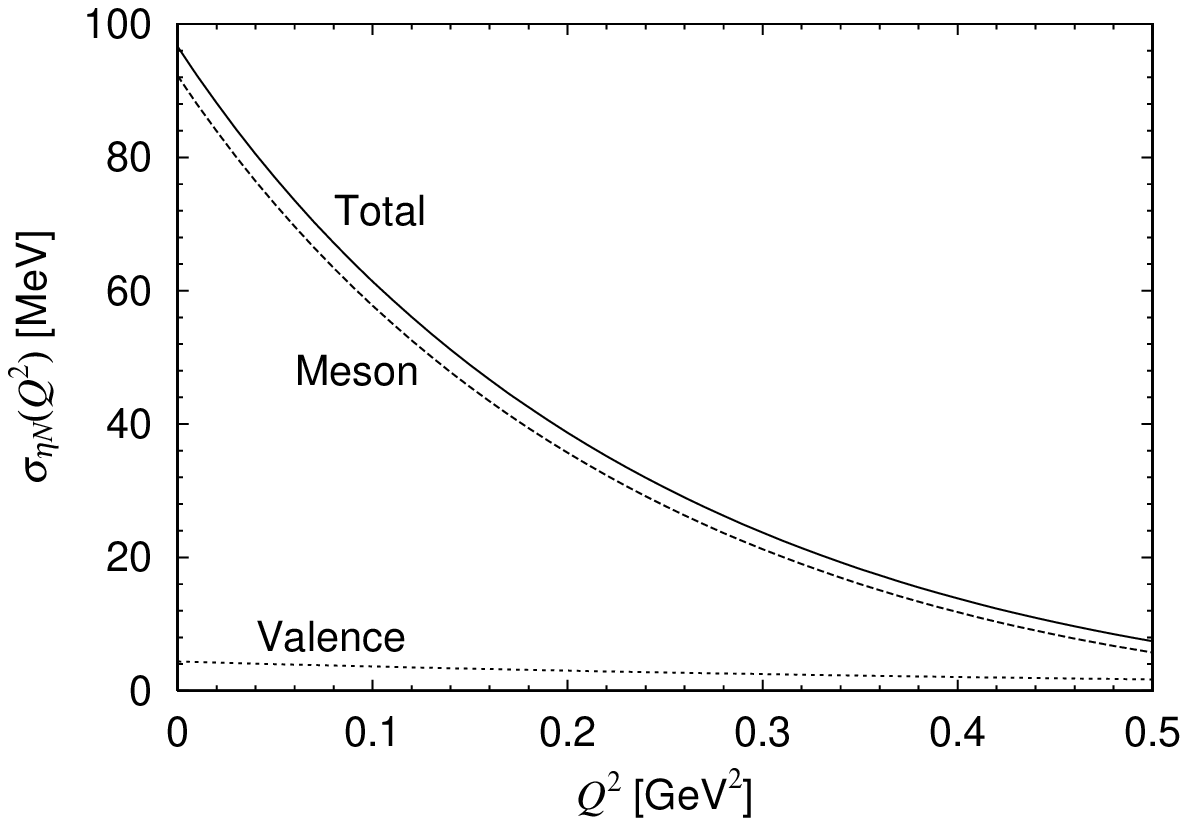,height=10cm}}
\end{figure}

\vspace*{2cm}

\centerline{\bf Fig.10}


\begin{thebibliography}{99}
%
\bibitem{Gasser:1980sb}
J.~Gasser,
Annals Phys.\  {\bf 136}, 62 (1981).
%
%
\bibitem{Reya:gk}
E.~Reya,
Rev.\ Mod.\ Phys.\  {\bf 46}, 545 (1974).
%
%
\bibitem{Jaffe:1979rq}
R.~L.~Jaffe,
Phys.\ Rev.\ D {\bf 21}, 3215 (1980).
%
%
\bibitem{Gasser:2000wv}
J.~Gasser and M.~E.~Sainio,
arXiv:hep-ph/0002283; 
M.~E.~Sainio,
PiN Newslett.\  {\bf 16}, 138 (2002)
[arXiv:hep-ph/0110413]. 
%
%
\bibitem{Schweitzer:2003sb}
P.~Schweitzer,
arXiv:hep-ph/0307336.
%
%
\bibitem{Gasser:1990ce}
J.~Gasser, H.~Leutwyler and M.~E.~Sainio,
Phys.\ Lett.\ B {\bf 253}, 252 (1991); 
Phys.\ Lett.\ B {\bf 253}, 260 (1991).
%
%
%
%
\bibitem{Koch:pu}
R.~Koch,
Z.\ Phys.\ C {\bf 15}, 161 (1982).
%
%
\bibitem{Becher:1999he}
T.~Becher and H.~Leutwyler,
Eur.\ Phys.\ J.\ C {\bf 9}, 643 (1999)
[arXiv:hep-ph/9901384].
%
%
\bibitem{Fuchs:2003kq}
T.~Fuchs, J.~Gegelia and S.~Scherer,
arXiv:hep-ph/0309234.
%
%
\bibitem{Borasoy:2002hz}
B.~Borasoy, R.~Lewis and P.~P.~Ouimet,
Phys.\ Rev.\ D {\bf 65}, 114023 (2002)
[arXiv:hep-ph/0203199].
%
%
\bibitem{Kaufmann:dd}
W.~B.~Kaufmann and G.~E.~Hite,
Phys.\ Rev.\ C {\bf 60}, 055204 (1999).
%
%
\bibitem{Olsson:1999jt}
M.~G.~Olsson,
Phys.\ Lett.\ B {\bf 482}, 50 (2000)
[arXiv:hep-ph/0001203].
%
%
\bibitem{Pavan:2001wz}
M.~M.~Pavan, I.~I.~Strakovsky, R.~L.~Workman and R.~A.~Arndt,
PiN Newslett.\  {\bf 16}, 110 (2002)
[arXiv:hep-ph/0111066].
%
%
\bibitem{Stahov:2002vs}
J.~Stahov,
arXiv:hep-ph/0206041.
%
%
\bibitem{Lyubovitskij:2000sf}
V.~E.~Lyubovitskij, T.~Gutsche, A.~Faessler and E.~G.~Drukarev,
Phys.\ Rev.\ D {\bf 63}, 054026 (2001)
[arXiv:hep-ph/0009341].
%
%
\bibitem{Lyubovitskij:2001nm}
V.~E.~Lyubovitskij, T.~Gutsche and A.~Faessler,
Phys.\ Rev.\ C {\bf 64}, 065203 (2001)
[arXiv:hep-ph/0105043].
%
%
\bibitem{Pumsa-ard:2003yh}
K.~Pumsa-ard, V.~E.~Lyubovitskij, T.~Gutsche, A.~Faessler and S.~Cheedket,
Phys.\ Rev.\ C {\bf 68}, 015205 (2003)
[arXiv:hep-ph/0304033].
%
%
\bibitem{Lyubovitskij:2001fv}
V.~E.~Lyubovitskij, T.~Gutsche, A.~Faessler and R.~Vinh Mau,
Phys.\ Lett.\ B {\bf 520}, 204 (2001) 
[arXiv:hep-ph/0108134]; 
%
%
Phys.\ Rev.\ C {\bf 65}, 025202 (2002)
[arXiv:hep-ph/0109213].
%
%
\bibitem{Simkovic:2001fy}
F.~Simkovic, V.~E.~Lyubovitskij, T.~Gutsche, A.~Faessler and S.~Kovalenko,
Phys.\ Lett.\ B {\bf 544}, 121 (2002)
[arXiv:hep-ph/0112277].
%
%
\bibitem{Lyubovitskij:2002ng}
V.~E.~Lyubovitskij, P.~Wang, T.~Gutsche and A.~Faessler,
Phys.\ Rev.\ C {\bf 66}, 055204 (2002)
[arXiv:hep-ph/0207225].
%
%
\bibitem{Khosonthongkee}K.~Khosonthongkee {\it et al.}, in preparation. 
%
%
\bibitem{Gasser:1987rb}
J.~Gasser, M.~E.~Sainio and A.~Svarc,
Nucl.\ Phys.\ B {\bf 307}, 779 (1988).
%
%
\bibitem{Gasser:1982ap}
J.~Gasser and H.~Leutwyler,
Phys.\ Rept.\  {\bf 87}, 77 (1982); 
Nucl.\ Phys.\ B {\bf 250}, 465 (1985).
%
%
\bibitem{Procura:2003ig}
M.~Procura, T.~R.~Hemmert and W.~Weise,
arXiv:hep-lat/0309020.
%
%
\bibitem{Dong:1995ec}
S.~J.~Dong, J.~F.~Lagae and K.~F.~Liu,
Phys.\ Rev.\ D {\bf 54}, 5496 (1996)
[arXiv:hep-ph/9602259].
%
%
\bibitem{Kim:1995hu}
H.~C.~Kim, A.~Blotz, C.~Schneider and K.~Goeke,
Nucl.\ Phys.\ A {\bf 596}, 415 (1996)
[arXiv:hep-ph/9508299].
%
%
\bibitem{Lee:1994jj}
C.~H.~Lee, G.~E.~Brown, D.~P.~Min and M.~Rho,
Nucl.\ Phys.\ A {\bf 585}, 401 (1995)
[arXiv:hep-ph/9406311].
%
%
\cite{Borasoy:1998uu}
\bibitem{Borasoy:1998uu}
B.~Borasoy,
Eur.\ Phys.\ J.\ C {\bf 8}, 121 (1999)
[arXiv:hep-ph/9807453].
%
%
\bibitem{Curceanu:2000dc}
C.~Curceanu {\it et al.} [DEAR Collaboration],
Nucl.\ Phys.\ A {\bf 691}, 278 (2001); 
R.~Baldini {\it et al.}  [DEAR collaboration],
LNF-95-055-IR
\end{thebibliography}
\end{document}